\documentclass[12pt]{article} 

\usepackage{packages}
\usepackage{allcommands}
\begin{document}
\justifying
\doublespacing

\begin{flushright}
Version dated: \today\\
\end{flushright}

\bigskip
\medskip
\begin{center}

\noindent{\Large \bf
Non-Linear Drivers of Population Dynamics: \\ 
a Nonparametric Coalescent Approach 
}
\bigskip

\noindent{\normalsize \sc
Filippo Monti$^{1}$, Nuno R.~Faria$^{2}$, Xiang Ji$^{3}$, Philippe Lemey$^{4}$, \\
Moritz U.G.~Kraemer$^{5,6}$, Marc A.~Suchard$^{1,7,8}$} \\
\singlespacing
\bigskip
\noindent {\small
  \it $^1$ 
  Department of Biostatistics, Jonathan and Karin Fielding School of Public Health, University of California Los Angeles, Los Angeles, CA, USA \\
  \it $^2$ 
  Department of Infectious Disease Epidemiology, School of Public Health, Imperial College London, London, UK \\
  \it $^3$ 
  Department of Statistics, Iowa State University, Ames, Iowa, USA \\
  \it $^4$ 
  Department of Microbiology, Immunology and Transplantation, Rega Institute, KU Leuven, Leuven, Belgium \\
  \it $^{5}$ 
  Department of Biology, University of Oxford, Oxford, UK \\
  \it $^{6}$ 
  Pandemic Sciences Institute, University of Oxford, Oxford, UK \\
  \it $^{7}$ 
  Department of Biomathematics, David Geffen School of Medicine at UCLA, University of California Los Angeles, Los Angeles, CA, USA \\
  \it $^{8}$ 
  Department of Human Genetics, David Geffen School of Medicine at UCLA, University of California Los Angeles, Los Angeles, CA, USA
} \\

\end{center}
\medskip
\noindent{\bf Corresponding author:} Marc A.~Suchard, Departments of Biostatistics, Biomathematics, and Human Genetics,
University of California Los Angeles, 695 Charles E.~Young Dr., South,
Los Angeles, CA 90095-7088, USA; E-mail: \url{msuchard@ucla.edu}

\vspace{1in}

\clearpage

\doublespacing
\paragraph{Abstract}
Effective population size ($N_e(t)$) is a fundamental parameter in population genetics and phylodynamics that quantifies genetic diversity and reveals demographic history. 
Coalescent-based methods enable the inference of $N_e(t)$ trajectories through time from time-scaled phylogenies reconstructed from molecular sequence data. 
Understanding the ecological and environmental drivers of population dynamics requires linking $N_e(t)$ to external data such as climate or epidemiological variables. 
Existing approaches typically impose log-linear relationships between covariates and $N_e(t)$, which may fail to capture complex biological processes and can introduce bias when the true relationship is nonlinear.
We present a flexible Bayesian framework that integrates covariates into coalescent models with piecewise-constant $N_e(t)$ through a Gaussian process (GP) prior. 
The GP, a distribution over functions controlled by a kernel with data-driven hyperparameters, naturally accommodates nonlinear covariate effects without restrictive parametric assumptions. 
This formulation improves estimation of covariate-$N_e(t)$ relationships, mitigates bias when associations are nonlinear, and yields interpretable uncertainty quantification that varies across the covariate space. 
To balance global covariate-driven patterns with local temporal dynamics, we couple the GP prior with a Gaussian Markov random field that enforces smoothness in $N_e(t)$ trajectories. 
Efficient inference is achieved via Hamiltonian Monte Carlo over the high-dimensional latent field.
Through simulation studies and three empirical applications—yellow fever virus dynamics in Brazil (2016–2018), late-Quaternary musk ox demography, and HIV-1 CRF02\_AG evolution in Cameroon—we demonstrate that our method both confirms linear relationships where appropriate and reveals nonlinear covariate effects that would otherwise be missed or mischaracterized. 
This framework advances phylodynamic inference by enabling more accurate and biologically realistic modeling of how environmental and epidemiological factors shape population size through time.

\vspace{1cm}


\clearpage

\bibliographystyle{apalike}
\section{Introduction}

\paragraph{The coalescent}
The \textbf{coalescent model}, introduced by \citet{kingman_coalescent_1982}, describes the genealogical process of a sample of genetic lineages under neutral evolution. 
Since its inception, numerous extensions have been developed to accommodate more complex demographic and evolutionary scenarios.  

\paragraph{The effective population size} 
A key parameter in this model is the \textbf{effective population size} ($\coEPSConstant$), which represents the size of an idealized population that would experience the same level of genetic drift as the actual population under study. 
Unlike the census population size, which counts all individuals, the effective population size accounts for factors such as variation in reproductive success, overlapping generations, and fluctuating population sizes, which affect the rate at which lineages coalesce when traced backward in time \citep{hudson_gene_1990}. 
In phylogenetics and population genetics, estimating the effective population size from genetic data is crucial for understanding past population dynamics, selection pressures, and demographic history. 

\paragraph{Time-varying effective population size} The original coalescent model \citep{kingman_coalescent_1982} assumes a constant population size, a simplification that is often unrealistic. 
To address this limitation, \citet{griffiths_sampling_1994} extended the model to allow for a time-varying effective population size, that is $\coEPSConstant = \coEPS{\coalescentTime{}}$. 
This generalization has led to a rich body of research developing inferential methods to reconstruct the effective population size dynamics.  


\paragraph{Parametric models}  
One approach to modeling time-varying effective population size is through parametric models, where the population size is assumed to follow a predefined functional form with a small number of parameters. 
A common example is exponential growth, as used in \citet{kuhner_maximum_1998} and \citet{drummond_estimating_2002}. 
The advantage of parametric models is that they provide a compact and interpretable representation of demographic history.
However, if the chosen functional form does not match the underlying trend, then the misspecification problem can lead to substantially biased inferences. 
Accurate results thus require a time-consuming model-selection process across diverse functional forms.
This motivated the development of semi- and non-parametric methods that accommodate more flexibility.

\paragraph{Piecewise-constant models} \citet{pybus_integrated_2000} initiated a new line of research by modeling the function $\coEPS{\coalescentTime{}}$ as piecewise-constant. 
Specifically, the authors allow $\coEPS{\coalescentTime{}}$ to change at each coalescent time, that is, every time a most recent common ancestor is found (\textit{classical skyline plot estimation}). 
This entails that the number of free parameters equals the number of observations; in other words, each effective population size estimate is inferred from a single observation, resulting in highly variable and noisy estimates.

To address these limitations, several solutions have been proposed.
\citet{strimmer_exploring_2001} reduced the number of free parameters of the aforementioned model by means of a model selection approach based on the Akaike Information Criterion (AIC) correction (\textit{generalized classical skyline plot estimation}).
Within a Bayesian framework, \citet{drummond_bayesian_2005} proposed pre-specifying a restricted total number of change points. To allow for more flexibility, \citet{opgen-rhein_inference_2005} assigned a prior to this number, relying on a reversible jump MCMC algorithm \citep{green_reversible_1995}. However, the results were highly sensitive to the prior choice.

\citet{minin_smooth_2008} proposed an alternative solution that anchored the change points to the coalescent times as in \citet{pybus_integrated_2000}, but assigned a (first-order) Gaussian Markov random field (GMRF) prior to the piecewise-constant levels of the population size function. Their main idea was to reduce overfitting  using a prior that penalizes the variability of $\coEPS{\coalescentTime{}}$ over short time windows (\textit{Bayesian Skyride}) by inducing a dependence between neighboring levels. 
Subsequently, \citet{gill_improving_2013} adopted the same GMRF-based approach, but fixed \textit{a priori} an equally-spaced grid of change points (\textit{Bayesian Skygrid}). 
This approach overcame the prior's dependence on the tree structure, which previously limited its application to analyses involving multiple trees. 
Building on this work, \citet{gill_understanding_2016} incorporated external covariates using a log-linear model for the GMRF mean of (log) effective population size levels. This allowed the introduction of additional information to further inform the inference process, inducing a dependence between non-neighboring levels, and, as a byproduct, providing insights into which factors could influence demographic fluctuations. 
Since $\coEPS{\coalescentTime{}}$ levels might be high in number and often exhibit strong posterior correlations, traditional random-walk MCMC methods tend to mix slowly. 
This motivated \citet{baele_hamiltonian_2020} to simulate their posterior distribution more efficiently using a gradient-based sampler: Hamiltonian Monte Carlo (HMC).

\paragraph{Non-parametric models} 
As an alternative to piecewise-constant models, \citet{palacios_gaussian_2013} introduced a fully non-parametric approach that models the effective population size $\coEPS{\coalescentTime{}}$ as a continuous function of time.
They placed a Gaussian process (GP) prior on a latent function of time, and linked it to the inverse population size through a logistic transformation, ensuring that $1/\coEPS{\coalescentTime{}}$ remains positive and bounded. 
This transformation allows them to treat the coalescent as a point process and to evaluate its likelihood using a thinning-based data augmentation scheme, which replaces intractable time integrals with a finite set of simulated latent points. While this strategy provides smooth, flexible trajectories for $\coEPS{\coalescentTime{}}$, it also introduces substantial computational overhead because each MCMC iteration must resample these latent points. 
Moreover, the framework is built entirely around time as the sole predictor, making it difficult to incorporate other covariates without redefining the thinning procedure and likelihood structure.

\paragraph{Limitations of available models}
Piecewise-constant models remain the most popular choice because they balance flexibility and computational feasibility.
Their main limitation is enforcing smoothness across $\coEPS{\coalescentTime{}}$ levels while retaining enough change points to capture meaningful demographic variation.
The GMRF prior of \citet{minin_smooth_2008} and \citet{gill_improving_2013} partly addresses this by linking adjacent $\coEPS{\coalescentTime{}}$ levels.
\citet{gill_understanding_2016} extended this idea by incorporating external covariates, inducing dependencies beyond neighboring levels.
However, their log-linear formulation constrains the relationship between $\coEPS{\coalescentTime{}}$ and covariates to a narrow functional class; when this assumption is violated, inference becomes biased and fails to exploit the full information provided by the covariates.

%

\paragraph{Proposed model}
We propose a hybrid approach that balances the computational efficiency of piecewise-constant models with the flexibility of non-parametric methods. 
Building on \citet{gill_understanding_2016}, we retain a GMRF prior with a fixed grid of change points but replace the log-linear mean structure with a Gaussian process prior defined over the covariates.
By assigning positive probability over a broad class of functions, the GP prior enables our model to capture the nonlinear effects of covariates and address the limitations of a log-linear model.
Besides, the GP prior is characterized by a parametrized \textit{kernel} that controls the covariance between the values of the random function evaluated at different input points. 
The choice of the kernel, along with its parameters, regulates the smoothness of the inferred function. 
This not only enables higher-order dependencies between non-neighboring levels of $\coEPS{\coalescentTime{}}$, but it also allows for a data-driven selection of smoothness by assigning a prior to the kernel's parameters. 
In contrast with \citet{palacios_gaussian_2013}, who used a time-based GP linked to a normalized transformation of $\coEPS{\coalescentTime{}}$
and relied on thinning to evaluate the coalescent likelihood, our model (1) preserves temporal coherence by using the GMRF prior, (2) integrates covariate information accommodating nonlinear effects using a GP, (3) avoids the computational burden of thinning-based inference.

\paragraph{Structure of the paper} In the following sections, we begin by formally introducing the general time-varying coalescent model (\ref{NPco.sec:generalTimeVaryingModel}), and then adapt it to a framework where the effective population size function is modeled as piecewise-constant (\ref{NPco.sec:piecewiseConstantModel}), distinguishing between contemporaneous and serial sampling. 
Subsequently, Section \ref{NPco.sec:coalescentBasedLikelihood} will draw some considerations on the use of the coalescent as a prior. 
Then, we present our choice for the prior on $\coEPS{\coalescentTime{}}$ levels (\ref{NPco.sec:modelingPopSizes}), compute the gradient and Hessian of the posterior distribution (\ref{NPco.sec:likelihoodGradientHessian}), and motivate their use within an HMC-based sampling scheme (\ref{NPco.sec:HMCSampling}).
Finally, to test the reliability and real-world applicability of our model, we apply our methods to synthetic and real data examples (\ref{NPco.sec:results}).

\section{Methods}
\paragraph{Overview and conventions.}
In Bayesian phylogenetics, the \textbf{coalescent process} serves as a \emph{tree prior}, that is, a probabilistic model describing how sampled genetic lineages merge when traced backward in time. 
Given molecular sequence data, this prior connects demographic parameters, i.e. the effective population sizes $\coEPSConstant$, to the shape and branch lengths of genealogical trees. 

We denote by $\phylogeny{}$ a rooted, binary \textbf{phylogeny} describing the ancestry of $\conTaxa$ sampled taxa.  
Since $\phylogeny{}$ is binary, it has $\conTaxa$ terminal nodes (tips) and $\conTaxa - 1$ internal nodes each representing a coalescent event where two lineages merge into a common ancestor.
The phylogeny $\phylogeny{}$ is fully characterized by two components:
(i) a \emph{topology} specifying which lineages coalesce, and 
(ii) a set of node times comprising both tip times and internal (coalescent) node times.

Throughout this work, we adopt a \emph{backward-time} convention: time increases from the present into the past, so smaller values correspond to more recent times.
To ease notation, we initially assume \emph{isochronous sampling}, meaning that all $\conTaxa$ tips are sampled contemporaneously at present time.
Under this assumption, we can set all tip times equal to zero, $\coalescentTime{0}=0$, and denote the internal node times as
\begin{align}
    0 < \coalescentTime{1} \leq \coalescentTime{2} \leq \cdots \leq \coalescentTime{\conTaxa-1},
\end{align}
where $\coalescentTime{\coTimeIndex}$ denotes the backward time of the $\coTimeIndex$-th coalescent event.
Each internal node represents a most recent common ancestor (MRCA) of two lineages existing at that time.
In Section~\ref{NPco.sec:piecewiseConstantModel}, we generalize to accommodate serial (heterochronous) sampling, where tips may be sampled at different times.


Conditional on the demographic parameters, the coalescent model specifies a probability density for the ordered set of coalescent times 
$(\coalescentTime{1}, \ldots, \coalescentTime{\conTaxa-1})$ that is compatible with $\phylogeny{}$, while we assume a uniform distribution over tree topologies.
Section~\ref{NPco.sec:coalescentBasedLikelihood} provides a construction of a probability distribution over labeled trees by extending the coalescent process to account explicitly for the identities of the lineages that coalesce over time.

\subsection{General Time-Varying Model} \label{NPco.sec:generalTimeVaryingModel}
We now formalize the coalescent process under the assumption that the effective population size may change over time, that is, $\coEPSConstant = \coEPS{\coalescentTime{}}$. 
Under isochronous sampling, the ordered sequence $\coalescentTime{1},\ldots,\coalescentTime{\conTaxa-1}$ lists the \textbf{coalescent times} (in backward time), where $\coalescentTime{\coTimeIndex}$ is the time of the $\coTimeIndex$-th coalescent event. 
Let $\conLineages{\coTimeIndex}$ be the number of extant lineages during the interval $[\coalescentTime{\coTimeIndex-1},\coalescentTime{\coTimeIndex})$; with a single coalescent event at $\coalescentTime{\coTimeIndex}$ we have the recursion $\conLineages{\coTimeIndex}=\conLineages{\coTimeIndex-1}-1=K + 1 - k$.

Fixing the time-varying effective population size function $\coEPS{\cdot}$, the conditional probability density that the $\coTimeIndex$-th coalescent event occurs at $\coalescentTime{\coTimeIndex}$ given $\coalescentTime{\coTimeIndex-1}$ is \citep{griffiths_sampling_1994}:
\begin{align}
\condprob{\coalescentTime{\coTimeIndex}}{\coalescentTime{\coTimeIndex-1}, \coEPS{\cdot}}
= \binom{\conLineages{\coTimeIndex}}{2}\frac{1}{\coEPS{\coalescentTime{\coTimeIndex}}}
\exp\!\left\{-\binom{\conLineages{\coTimeIndex}}{2}\int_{\coalescentTime{\coTimeIndex-1}}^{\coalescentTime{\coTimeIndex}} \frac{1}{\coEPS{\tau}}\,d\tau\right\}.
\end{align}
Here, $\binom{\conLineages{\coTimeIndex}}{2}\coEPS{\coalescentTime{\coTimeIndex}}^{-1}$ is the instantaneous probability, or \emph{hazard}, that two out of the $\conLineages{\coTimeIndex}$ nodes coalesce at time $\coalescentTime{\coTimeIndex}$, while the exponential term is the \emph{survival} probability of no coalescence in $(\coalescentTime{\coTimeIndex-1},\coalescentTime{\coTimeIndex})$. Multiplying over events yields the joint density of coalescent times:
\begin{align}
\condprob{\coalescentTime{1},\ldots,\coalescentTime{\conTaxa-1}}{\coEPS{\cdot}}
= \bracel\prod_{\coTimeIndex=1}^{\conTaxa-1}\binom{\conLineages{\coTimeIndex}}{2}\frac{1}{\coEPS{\coalescentTime{\coTimeIndex}}}\, \bracer
\exp\!\left\{-\sum_{\coTimeIndex=1}^{\conTaxa-1}\binom{\conLineages{\coTimeIndex}}{2}\int_{\coalescentTime{\coTimeIndex-1}}^{\coalescentTime{\coTimeIndex}}\frac{1}{\coEPS{\tau}}\,d\tau\right\}.
\end{align}

\subsection{Piecewise-Constant Model} \label{NPco.sec:piecewiseConstantModel}
A common semi-parametric approach to modeling a time-varying effective population size function $\coEPS{\coalescentTime{}}$ is to assume it is \textbf{piecewise-constant} over a fixed grid of time intervals.  
Let $\coGridPoint{0}, \coGridPoint{1}, \ldots, \coGridPoint{\conGridPoints-1}$ denote a sequence of \textbf{grid points} that partition the time axis into $\conGridPoints$ intervals starting at $\coGridPoint{0} = \coalescentTime{0}$.  
These are treated as \emph{exogenous}, meaning they are specified independently of the realized coalescent times. 
A simple and effective choice is to use an equally spaced grid, though any monotone sequence may be employed.
The effective population size is then defined as
\begin{align}
    \coEPS{\coalescentTime{}} = \coPiecewiseConstantLevels{\coGridPointIndex}
    \quad \text{for } \coGridPoint{\coGridPointIndex-1} < \coalescentTime{} \leq \coGridPoint{\coGridPointIndex},
    \qquad \coGridPointIndex = 1, 2, \ldots, \conGridPoints -1 .
\end{align}
Because the root time may extend beyond the last predefined grid point, we append an additional point $\coGridPoint{\conGridPoints}$ placed sufficiently far in the past and denote the population size between $\coGridPoint{\conGridPoints-1}$ and $\coGridPoint{\conGridPoints}$ by $\coPiecewiseConstantLevels{\conGridPoints}$. 
The length of this interval has no inferential role; in practice, we simply let the final grid interval extend up to the realized root height.


\paragraph{Isochronous data} For clarity, we first derive the joint density assuming all $\conTaxa$ taxa are sampled contemporaneously (isochronous case). 
Let $\conCoalescentEventsGridPoints{\coGridPointIndex}$ be the number of coalescent events (internal nodes) between grid points $\coGridPoint{\coGridPointIndex-1}$ and $\coGridPoint{\coGridPointIndex}$.  
Moreover, let 
\begin{align}
    \coalescentTimeVector{\coGridPointIndex} = 
    \left(\coalescentTime{\coGridPointIndex,1}, \coalescentTime{\coGridPointIndex,2}, \ldots, 
\coalescentTime{\coGridPointIndex,\conCoalescentEventsGridPoints{\coGridPointIndex}} \right)
\end{align}
 be the sequence of coalescent events between time $\coGridPoint{\coGridPointIndex-1}$ and $\coGridPoint{\coGridPointIndex}$. 
We further collect these terms into the vector $\coalescentTimeVector{} = (\coalescentTimeVector{1}, \coalescentTimeVector{2}, \dots, \coalescentTimeVector{\conGridPoints})$.
Finally, let $\conLineagesGridPoints{\coGridPointIndex, \coTimeIndex}$ be the number of non-coalesced nodes immediately before time $\coalescentTime{\coGridPointIndex, \coTimeIndex}$.

Following the same hazard–survival reasoning as in Section~\ref{NPco.sec:generalTimeVaryingModel}, the joint probability density of all coalescent times can be written as

\begin{align}
\condprob{\coalescentTimeVector{}}{\coPiecewiseConstantLevelsVector}
&= \Biggl\{\prod_{\coGridPointIndex=1}^{\conGridPoints}\prod_{\coTimeIndex=1}^{\conCoalescentEventsGridPoints{\coGridPointIndex}}
\left[\binom{\conLineagesGridPoints{\coGridPointIndex,\coTimeIndex}}{2}\frac{1}{\coPiecewiseConstantLevels{\coGridPointIndex}}\right]\Biggr\}
\exp \Biggl\{-\sum_{\coGridPointIndex=1}^{\conGridPoints}\sum_{\coTimeIndex=1}^{\conCoalescentEventsGridPoints{\coGridPointIndex}}
\binom{\conLineagesGridPoints{\coGridPointIndex,\coTimeIndex}}{2}\frac{\coalescentTime{\coGridPointIndex,\coTimeIndex}-\coalescentTime{\coGridPointIndex,\coTimeIndex-1}}{\coPiecewiseConstantLevels{\coGridPointIndex}}\Biggr\} \nonumber\\
&=\Biggl\{\prod_{\coGridPointIndex=1}^{\conGridPoints}\Bigl[  \frac{1}{\coPiecewiseConstantLevels{\coGridPointIndex}} \Bigr]^{\conCoalescentEventsGridPoints{\coGridPointIndex}}
\prod_{\coTimeIndex=1}^{\conCoalescentEventsGridPoints{\coGridPointIndex}}\binom{\conLineagesGridPoints{\coGridPointIndex,\coTimeIndex}}{2}\Biggr\}
\exp \Biggl\{-\sum_{\coGridPointIndex=1}^{\conGridPoints}\frac{1}{\coPiecewiseConstantLevels{\coGridPointIndex}}
\sum_{\coTimeIndex=1}^{\conCoalescentEventsGridPoints{\coGridPointIndex}}
\binom{\conLineagesGridPoints{\coGridPointIndex,\coTimeIndex}}{2}\bigl(\coalescentTime{\coGridPointIndex,\coTimeIndex}-\coalescentTime{\coGridPointIndex,\coTimeIndex-1}\bigr)\Biggr\}, 
\end{align}
where $\binom{0}{2} = 0$ and, for notational convenience, we define the boundary values
$\coalescentTime{\coGridPointIndex,0} := \coGridPoint{\coGridPointIndex-1}$ and
$\coalescentTime{\coGridPointIndex,\conCoalescentEventsGridPoints{\coGridPointIndex}+1} := \coGridPoint{\coGridPointIndex}$,
so that the inner summation in the exponent can be written in a unified form.

\paragraph{Heterochronous data} To account for heterogeneity in tip-sampling times, we need to redefine $\coalescentTime{\coGridPointIndex, \coTimeIndex}$ to be either a \textit{coalescent time} or a \textit{tip sampling date}, and $\conLineagesGridPoints{\coGridPointIndex,\coTimeIndex}$ to be updated accordingly.
Indeed, the order of tip sampling dates impacts the number of lineages that can coalesce in a determined time window. 
For example, suppose $\coalescentTime{\coGridPointIndex, \coTimeIndex}$ coincides with a sampling date (and not a coalescent event), then the probability of a subsequent coalescent event increases since more nodes can share a common ancestor. 

We use $\conCoalescentEventsGridPoints{\coGridPointIndex}$ to count the number of coalescent events in  $(\coGridPoint{\coGridPointIndex-1}, \coGridPoint{\coGridPointIndex}]$, while we denote $\conSamplingTipsDatesGridPoints{\coGridPointIndex}$ the number of tip sampling dates in the same period. Moreover, we introduce the sets $\left\{ \cosetCoalescentTimes{\coGridPointIndex}\right\}$ where $\cosetCoalescentTimes{\coGridPointIndex}$ collects the indices of all the coalescent events that occurred between grid points $\coGridPoint{\coGridPointIndex - 1}$ and $\coGridPoint{\coGridPointIndex}$.
Given this new notation, the likelihood becomes: 
\begin{align} \label{NPco.eq:coheteroPiecewiseLikelihood}
    \condprob{\coalescentTimeVector{}}{\coPiecewiseConstantLevelsVector}
    &=  \bracel
    \prod_{\coGridPointIndex=1}^{\conGridPoints} \left[\frac{1}{\coPiecewiseConstantLevels{\coGridPointIndex}}  \right]^{\conCoalescentEventsGridPoints{\coGridPointIndex}}
    \prod_{\coTimeIndex \in \cosetCoalescentTimes{\coGridPointIndex}} \binom{\conLineagesGridPoints{\coGridPointIndex,\coTimeIndex}}{2} \bracer
    \myExp{-\sum_{\coGridPointIndex=1}^{\conGridPoints}\frac{1}{\coPiecewiseConstantLevels{\coGridPointIndex}}  \sum_{\coTimeIndex=1}^{\conCoalescentEventsGridPoints{\coGridPointIndex} + \conSamplingTipsDatesGridPoints{\coGridPointIndex}} \binom{\conLineagesGridPoints{\coGridPointIndex,\coTimeIndex}}{2} (\coalescentTime{\coGridPointIndex,\coTimeIndex} - \coalescentTime{\coGridPointIndex,\coTimeIndex-1})}.
\end{align}
The only notable difference is that the internal summation in the exponent has become more granular: as mentioned above, this accounts for the increase in the probability of a coalescent event when new samples are collected.

\subsection{Coalescent-Based Likelihood(s)} \label{NPco.sec:coalescentBasedLikelihood}
The coalescent formulation specified in Equation~\eqref{NPco.eq:coheteroPiecewiseLikelihood} places a distribution on \emph{when} events occur (the internal node times) but is agnostic about \emph{which} lineages merge. 
It therefore defines a density over equivalence classes of trees that share the same ordered coalescent times. 
To obtain a probability distribution over fully labeled trees, this density must be extended to account for the identities of the lineages involved in each merger.
Equivalently, for a given ordered sequence of coalescent times, the probability mass is distributed uniformly across all labeled trees compatible with those times.

To count such trees, we fix an ordered sequence of coalescent times and the corresponding numbers of extant lineages immediately prior to each event.
At each coalescent time $\coalescentTime{\coGridPointIndex,\coTimeIndex}$, any unordered pair of the
$\conLineagesGridPoints{\coGridPointIndex,\coTimeIndex}$ extant lineages may merge, yielding
$\binom{\conLineagesGridPoints{\coGridPointIndex,\coTimeIndex}}{2}$ possible choices.
Because each choice uniquely determines the subsequent lineage configuration, the total number of
distinct labeled trees compatible with the given ordered coalescent times is
\begin{align} \label{co.eq:coLikelihoodNormalizingConstant}
\prod_{\coGridPointIndex=1}^{\conGridPoints}\;\prod_{\coTimeIndex\in\cosetCoalescentTimes{\coGridPointIndex}}
\binom{\conLineagesGridPoints{\coGridPointIndex,\coTimeIndex}}{2}.
\end{align}
Dividing \eqref{NPco.eq:coheteroPiecewiseLikelihood} by this quantity generates a properly normalized density on labeled trees.

\paragraph{Multilocus generalization} If there are more genealogies $\phylogenies{1}, \dots \phylogenies{\nPhylogenies}$ (e.g. associated with different loci in a genome), and assuming they are conditionally independent given the effective population sizes, then we can write their joint distribution as follows:
\begin{align}
    \condprob{\phylogeniesVector}{\coEPS{\cdot}} = \prod_{i=1}^{\nPhylogenies} \condprob{\phylogenies{i}}{\coEPS{\cdot}}.
\end{align}
All the methods presented hereafter can be easily generalized to accommodate additional genealogies.

\subsection{Modeling the Population Size} \label{NPco.sec:modelingPopSizes}
To avoid complications arising from the positivity constraint on each effective population size, we focus on its log-transformation:
\begin{align}
\piecewiseConstantAuxiliaryVariable{\coGridPointIndex} = \log{\coPiecewiseConstantLevels{\coGridPointIndex}}.
\end{align}
We then model the sequence of levels $\piecewiseConstantAuxiliaryVariable{\coGridPointIndex}$ leveraging two types of information: (1) chronological order, following the idea that we do not expect sudden changes in $\coEPS{\coalescentTime{}}$ over short periods of time; (2) external covariates, further informing the inferential process, contributing to limiting sudden variation, and shedding light on which factor might drive demographic fluctuations. 

In Section \ref{co.sec:GaussianProcess} we focus on modeling $\piecewiseConstantAuxiliaryVariable{\coGridPointIndex}$ directly using covariates based on a GP prior. 
Then, in Section \ref{co.sec:gmrfGP} we leverage the chronological information to model the GP error using a (first order) GMRF prior, or put another way, we use the GP to model the GMRF mean. This joint approach balances the \textit{global} dependence across all levels induced by the covariates with a \textit{local} time-dependence.

\subsubsection{Global Covariate-Dependence: Gaussian Process (GP) Prior} \label{co.sec:GaussianProcess}
Suppose that at each grid point $\coGridPointIndex$ we observe a set of $\conCovariates$ covariates $\coCovariateVector{\coGridPointIndex} = 
(\coCovariatesVariable{1 \coGridPointIndex },\coCovariatesVariable{2 \coGridPointIndex } \dots, \coCovariatesVariable{\conCovariates \coGridPointIndex})$; for example, $\coCovariatesVariable{i \coGridPointIndex}$ could be the temperature at time $\coGridPointIndex$.
The choice of these covariates can be motivated by two main objectives: either to guide the inference process—for example, when we have prior knowledge of a variable's correlation with the demographic history of a pathogen—or to directly investigate their association with $\coEPS{\coalescentTime{}}$.
%
%
We can then set the log effective population sizes equal to a random (real-valued) function of these covariates plus an error as follows: 
\begin{align}
\piecewiseConstantAuxiliaryVariable{\coGridPointIndex} = \log{\coPiecewiseConstantLevels{\coGridPointIndex}} = 
     \coGpFunctions{}{\coCovariateVector{\coGridPointIndex}} + \coGPErrorVariable{\coGridPointIndex},
\end{align}
where the errors $\coGPErrorVariable{\coGridPointIndex}$ are assumed jointly distributed as a multivariate normal with mean $\boldsymbol{0}$ and precision matrix $\coGmrfPrecisionMatrix^*$, that is, $\coGPErrorVector = (\coGPErrorVariable{1}, \coGPErrorVariable{2}, \dots, \coGPErrorVariable{\conGridPoints}) \sim \mathcal{MVN}(\boldsymbol{0}, (\coGmrfPrecisionMatrix^*)^{-1})$.

Following \citet{minin_smooth_2008}, \citet{gill_improving_2013} assumed a \textit{constant} function $\coGpFunctions{}{\coCovariateVector{\coGridPointIndex}}\equiv \beta_{0}$ and restricted the precision $\coGmrfPrecisionMatrix^*$ to the structure characteristic of a (first-order) GMRF prior. 
We will discuss and motivate this choice in the next section. 
\citet{gill_understanding_2016} later extended this framework assuming a (log-)linear model for the covariates, that is, fixing the covariate function equal to $\coGpFunctions{}{\coCovariateVector{\coGridPointIndex}} = \sum_{\coCovariateIndex=1}^{\conCovariates} \coCovariatesVariable{    \coCovariateIndex \coGridPointIndex} \beta_{\coCovariateIndex}$ and assigning iid normal priors to the linear coefficients $\beta_{\coCovariateIndex}$.
While convenient for interpretation, the log-linear assumption severely restricts the range of candidate functional forms. This limitation can both prevent a full utilization of covariate information and lead to a failure to capture nonlinear relationships, unless one is willing to perform model selection over a series of covariate transformations.

To overcome this restriction, we place a \textbf{Gaussian process} (\textbf{GP}) prior on the relationship between the coalescent function and the covariates, thereby substantially enhancing the model’s capacity to capture complex, nonlinear dependencies.
For expositional convenience, we assume that covariate effects combine additively, that is $\coGpFunctions{}{\coCovariateVector{\coGridPointIndex}} = \sum_{\coCovariateIndex=1}^{\conCovariates}\coGpFunctions{\coCovariateIndex}{\coCovariatesVariable{\coCovariateIndex\coGridPointIndex}}$ (even though this is not necessary as explained below). 
Then, we can assign independent GP priors to each of the functions $\coGpFunctions{\coCovariateIndex}{}$ such that 
\begin{align} \label{co.Eq:usingGps}
\piecewiseConstantAuxiliaryVariable{\coGridPointIndex} = \log{\coPiecewiseConstantLevels{\coGridPointIndex}} = 
    \sum_{\coCovariateIndex = 1}^{\conCovariates} \coGpFunctions{\coCovariateIndex}{\coCovariatesVariable{\coCovariateIndex\coGridPointIndex}} + \epsilon_\coGridPointIndex\quad \text{with} \quad \coGpFunctions{\coCovariateIndex}{\cdot} \overset{\text{ind}}{\sim}\coGP{0}{\coGpKernels{\coGpKernelsParameter_{\coCovariateIndex}}{\cdot,\cdot} } ,
\end{align}
where $\coGpKernels{\coGpKernelsParameter_{\coCovariateIndex}}{\cdot, \cdot}$ is the GP kernel that measures the covariance of the function values at each pair of covariate points, and $\coGpKernelsParameter_{\coCovariateIndex}$ are its hyperparameters.

\paragraph{Background on Gaussian processes}
A Gaussian process (GP) is a flexible prior for unknown functions $f$. 
It can be viewed as an infinite-dimensional generalization of a multivariate normal distribution (MVN) and, importantly, it is specified such that for any finite set of $m$ inputs 
$x_{1}, \ldots, x_{m}$, the corresponding function values 
$(f(x_{1}), \ldots, f(x_{m}))$ 
are jointly Gaussian. 
A GP is fully defined by two components: a mean function $\mu(\cdot)$ and a covariance function (kernel) $\coGpKernels{}{\cdot,\cdot}$. 
The mean function describes the expected function value at each input, while the kernel determines the covariance between the function values at different points. 
Under these definitions, the vector $(f(x_{1}),\allowbreak \ldots, 
\allowbreak f(x_{m}))$ is distributed as a multivariate normal distribution with mean $(\mu(x_{1}), \allowbreak \ldots, \allowbreak \mu(x_{m}))$ and covariance matrix whose $ij$-th element is $\coGpKernels{}{x_i,x_j}$.

A GP prior offers great flexibility in modeling nonlinear relationships without assuming a specific functional form. 
A key element of a GP model is the choice of its kernel. 
Indeed, the kernel provides a measure of similarity between inputs and dictates the smoothness and shape of the functions drawn from the process. 
For example, the squared exponential kernel favors smoother functions, while the Mat\'ern kernel tends to accommodate more erratic behavior \citep{rasmussen_gaussian_2006}. 

\paragraph{Kernel choice} 
In our context we expect a mostly smooth function with at most two or three inflection points. Indeed, it is unlikely that the effect of a covariate changes sign multiple times.
Therefore, the squared exponential kernel seems the most appropriate choice. 
This kernel is characterized by two positive hyperparameters, the marginal scale ($\sigma^2$) and the length-scale ($\ell$), and it is defined as follows: 
\begin{align}\label{co:EqSquaredExpKernel}
    \coGpKernels{}{\coCovariatesVariable{i},\coCovariatesVariable{j}} = \sigma^2 \myExp{-\frac{(\coCovariatesVariable{i}-\coCovariatesVariable{j})^2}{2\ell^2}}.
\end{align}
The marginal scale controls the overall magnitude of the function's variation, while the length-scale determines the degree of regularity or how rapidly correlations decay with distance in the input space.
In particular, low values of $\ell$ correspond to more function oscillations, while increasing $\ell$ makes the function approach linearity. 
Within a Bayesian framework, we can assign priors to each of the hyperparameters, enabling a data-driven determination of their values.
We propose to use independent exponential priors with rate $1$ on each of the two kernel hyperparameters.
Moreover, in case the GP clearly overfits, that is, it displays unrealistically rapid oscillations given prior biological knowledge, we suggest constraining the length-scale by enforcing a strictly positive non-zero lower bound, thereby discouraging excessively short-range correlations.

\paragraph{Marginal distribution} Returning to Equation~\eqref{co.Eq:usingGps}, we can exploit the independence of the priors and the errors to write:
\begin{align}
    \piecewiseConstantAuxiliaryVariable{\coGridPointIndex} \sim \coN{\boldsymbol{0}}{ \coGpKernels{\coGpKernelsParameter}{\coCovariateVector{\coGridPointIndex}^{\phantom{\transpose}},\coCovariateVector{\coGridPointIndex}\transpose} + (\coGmrfPrecisionMatrix^*)^{-1}_{\coGridPointIndex\coGridPointIndex}},
\end{align}
where $\coGpKernels{\coGpKernelsParameter}{\coCovariateVector{\coGridPointIndex}^{\phantom{\transpose}},\coCovariateVector{\coGridPointIndex}\transpose} := \sum_{\coCovariateIndex = 1}^{\conCovariates} \coGpKernels{\coGpKernelsParameter_{\coCovariateIndex}}{\coCovariatesVariable{\coCovariateIndex\coGridPointIndex}^{\phantom{\transpose}},\coCovariatesVariable{\coCovariateIndex\coGridPointIndex}\transpose}$.
In vector form, we can collect all the log effective population size levels $\piecewiseConstantAuxiliaryVector = (\piecewiseConstantAuxiliaryVariable{1},  \piecewiseConstantAuxiliaryVariable{2}, \dots,  \piecewiseConstantAuxiliaryVariable{\conGridPoints})$, and express their joint distribution as a multivariate normal:
\begin{align}
    \piecewiseConstantAuxiliaryVector \sim \coMVN{\boldsymbol{0}}{\coGpKernelsMatrix{\coGpKernelsParameter}{\coCovariatesMatrix^{\phantom{\transpose}},\coCovariatesMatrix\transpose} + (\coGmrfPrecisionMatrix^*)^{-1}},
\end{align}
where $\coCovariatesMatrix = (\coCovariateVector{1}, \coCovariateVector{2}, \dots, \coCovariateVector{\conGridPoints}$) is a matrix with columns equal to the grid point-specific covariates, while $\coGpKernelsMatrix{\coGpKernelsParameter}{\coCovariatesMatrix^{\phantom{\transpose}},\coCovariatesMatrix\transpose}$ is a matrix with element $(\coGridPointIndex_i, \coGridPointIndex_j)$ equal to 
$\coGpKernels{\coGpKernelsParameter}{\coCovariateVector{\coGridPointIndex_i}^{\phantom{\transpose}},\coCovariateVector{\coGridPointIndex_j}\transpose}$.
This framework can be extended to include interactions between covariates, allowing for even more flexible inference.

\subsubsection{Modeling the Error Precision: Local Time-Dependence and GMRF}\label{co.sec:gmrfGP}

\paragraph{Error precision matrix $\coGmrfPrecisionMatrix^*$} An adequate choice for the structure of the errors $\coGPErrorVector$'s precision matrix $\coGmrfPrecisionMatrix^*$ is crucial. 
Indeed, choosing a completely unrestricted matrix $\coGmrfPrecisionMatrix^*$ introduces $\conGridPoints(\conGridPoints+1)/2$ additional parameters, dramatically increasing the parameter-space dimension and possibly leading to unidentifiability challenges or overfitting. 
A popular solution is to restrict $\coGmrfPrecisionMatrix^*$ to a diagonal matrix, that is, to assume the Gaussian errors $\coGPErrorVariable{\coGridPointIndex}$ are independent and possibly with a common variance. However, this approach has several drawbacks. 
First, the model becomes very sensitive to outliers among covariates and (``observed") function values. 
For example, $\coEPS{\coalescentTime{}}$ levels associated with covariate outliers would suffer higher variability as the GP would not have enough information to capture their value. 
This problem is further exacerbated when the covariate outliers are observed at the extremes of the time range. 
In GP inference, outlying \textit{observed} GP function values can significantly influence---and so bias---the posterior function, particularly with smooth kernels like the squared exponential. 
Even though, in our context, GP function values are the $\operatorname{log}\coEPS{\coalescentTime{}}$ levels which are \textit{not} observed directly, their posterior distribution is driven by the observed data through the coalescent prior bringing up a similar, though less-serious, issue.

Second, independent errors imply that, conditional on the covariates, there is no residual correlation between neighboring grid points. 
This means that the model does not account for any local temporal dependence that is not explained by the covariates. 
However, since the grid points are ordered chronologically, we would expect a residual correlation between $\coEPS{\coalescentTime{}}$ neighboring time points, suggesting the matrix $\coGmrfPrecisionMatrix^*$ should be more structured.



\paragraph{Gaussian Markov random field} Following \citet{minin_smooth_2008} and \citet{gill_improving_2013}, we model the errors using a \textit{first-order} GMRF inducing a direct co-dependence between adjacent errors. 
In particular, we assume that the vector $\coGPErrorVector$ has the density of a multivariate normal distribution with mean $\boldsymbol{0}$ and precision $ \coGmrfPrecisionMatrix^* = \coGmrfPrecision \coGmrfPrecisionMatrix$  such that:
\begin{align} \label{co.Eq:gmrfErrors}
    \prob{\coGPErrorVector} \propto \myExp{-\frac{\coGmrfPrecision}{2}\coGPErrorVector \transpose \coGmrfPrecisionMatrix\coGPErrorVector}  = \myExp{-\frac{\coGmrfPrecision}{2} \sum_{\coGridPointIndex=2}^{\conGridPoints}(\coGPErrorVariable{\coGridPointIndex}  - \coGPErrorVariable{\coGridPointIndex - 1} )^2 },
\end{align}
where $\coGmrfPrecision$ is the scalar precision parameter and $\coGmrfPrecisionMatrix$ is a tri-diagonal matrix with diagonal elements equal to $\coGmrfPrecisionMatrix_{ii}=2$ for $i = 2, \dots, \conGridPoints-1$ and $\coGmrfPrecisionMatrix_{1,1}=1=\coGmrfPrecisionMatrix_{\conGridPoints,\conGridPoints}$, and non-diagonal elements equal to $\coGmrfPrecisionMatrix_{i, i-1}= \coGmrfPrecisionMatrix_{i, i+1} = -1$. 
The structure of $\coGmrfPrecisionMatrix$ determines the main computational advantage of this approach: the GMRF prior can be evaluated in linear time.

The precision parameter $\coGmrfPrecision$ plays a crucial role in controlling the GMRF smoothness. 
A larger value of $\coGmrfPrecision$ indicates that the squared differences between consecutive values, $(\coGPErrorVariable{\coGridPointIndex} - \coGPErrorVariable{\coGridPointIndex-1})^2$, are penalized more heavily. 
This forces those values to be closer to each other, resulting in a smoother series of errors. 
Conversely, a smaller value of $\coGmrfPrecision$ allows for larger jumps between consecutive values, leading to a less smooth, or more jagged, series. 
The precision parameter thus acts as a hyperparameter that controls the degree of local temporal dependence, and we can estimate its value directly from the data.


\paragraph{Hyperprior on the precision parameter} 
We propose to assign an \emph{inverse-Gamma} prior with shape parameter $1$ and scale parameter $10$ to the GMRF precision parameter. 
This choice replaces overly diffuse alternatives used in previous works, such as a \emph{Gamma} prior with shape and scale parameters equal to $0.001$ and $1000$. 
Indeed, we found that in our context the \emph{Gamma} prior concentrates excessive mass near zero and yields unrealistically large marginal variances. 
The adopted specification provides a more balanced compromise, retaining weak informativeness while discouraging pathological under-smoothing of the latent process.

\paragraph{Alternative GMRF parametrization} The approach proposed in \citet{minin_smooth_2008} and \citet{gill_improving_2013}, and then extended by \citet{gill_understanding_2016} to incorporate covariates, used a different---though equivalent---parametrization that we also report here for completeness. 
Specifically, their ``GMRF-centric" approach models the levels $\piecewiseConstantAuxiliaryVariable{\coGridPointIndex}$ directly with a GMRF, whose mean is set to be a random function of the covariates, that is $\coGmrfMeanVariable{\coGridPointIndex} = \coGmrfMeanVariable{\coGridPointIndex}(\coCovariateVector{\coGridPointIndex})$. 
\citet{gill_understanding_2016} restricts such a function to a (log)-linear shape, while in this work, as mentioned above, we assign it a more flexible GP prior. 
The full hierarchical model can then be rewritten as: 
\begin{align} \label{co.eq:gmrfCentricLogPopDistribution}
 \piecewiseConstantAuxiliaryVector | \coCovariatesMatrix &\sim \mathcal{MVN}\left( \coGmrfMeanVector{}{\coCovariatesMatrix}, \frac{1}{\coGmrfPrecision}\coGmrfPrecisionMatrix^{-1}\right)   \\
\coGmrfMeanVector{}{\coCovariatesMatrix} & \sim \coMVN{\boldsymbol{0}}{\coGpKernelsMatrix{\coGpKernelsParameter}{\coCovariatesMatrix^{\phantom{\transpose}},\coCovariatesMatrix\transpose}}.
\end{align}
where $\coGmrfMeanVector{}{\coCovariatesMatrix}:= (\coGmrfMeanVariable{1}(\coCovariateVector{1}), \coGmrfMeanVariable{2}(\coCovariateVector{2}),  \dots, \coGmrfMeanVariable{\conGridPoints}(\coCovariateVector{\conGridPoints}))$ is the vector of GMRF means.



\subsection{Inference}

In this work, we use a Bayesian framework to infer both the evolutionary history of a population and its demographic changes over time. 
The core objective of phylogenetics is to reconstruct a phylogenetic tree ($\phylogenies{}$), which visualizes the evolutionary relationships among genetic sequences ($\mathbf{y}$).
It is customary to assume a continuous-time Markov chain (CTMC) parametrized by a rate matrix ($\coCtmcRateMatrix$) to model DNA substitutions and so to generate these sequences by simulating evolution along the tree. 
Typical phylogenetic models also assume that the rate of evolution may change over time or across sites. 
The current work does not affect these modeling choices, so the evolutionary rate will not be explicitly parametrized for notational simplicity.

The foundation of Bayesian inference relies on Bayes' theorem, which allows us to calculate the posterior distribution of the unknown parameters—e.g., the phylogenetic tree ($\phylogenies{}$) and the rate matrix ($\coCtmcRateMatrix$)—given the observed data ($\mathbf{y}$). 
This posterior distribution is proportional to the product of the data likelihood and the priors on the unknown parameters. 
In phylogenetics, the data likelihood, $\condprob{\mathbf{y}}{\phylogenies{}, \coCtmcRateMatrix}$, is equal to the probability of the observed sequences being generated by the CTMC along the tree. 
The choice of the prior for $\coCtmcRateMatrix$ is beyond the scope of this work and depends on the parametrization of the CTMC rate matrix \citep[see, e.g.,][]{kimura_simple_1980, jukes_evolution_1969}. 
A treatment of the prior on $\mathbf{Q}$ complementary to the current work—that is, using Gaussian processes with covariates—is provided in \citet{monti2025nonparametric}.

Conversely, our focus here is on the tree prior.
In particular, we adopt a coalescent process prior $\condprob{\phylogenies{}}{\coEPS{\coalescentTime{}}}$ with piecewise-constant population sizes (see Equation~\eqref{NPco.eq:coheteroPiecewiseLikelihood}). Then, we specify the prior over the population size levels $\condprob{\coEPS{\coalescentTime{}}}{\coCovariatesMatrix}$ leveraging temporal information and external covariates by combining GMRF and GP priors as explained in Section \ref{NPco.sec:modelingPopSizes}.
We can now combine all these elements to write down the posterior distribution of the unknown parameters $(\phylogenies{}, \coEPS{\coalescentTime{}}, \coCtmcRateMatrix)$ given the observed data $\mathbf{y}$ as follows:
\begin{align}
    \condprob{\phylogenies{}, \coEPS{\coalescentTime{}}, \coCtmcRateMatrix}{\mathbf{y}} 
    \propto \condprob{\mathbf{y}}{\phylogenies{}, \coCtmcRateMatrix} \prob{\coCtmcRateMatrix}
    \condprob{\phylogenies{}}{\coEPS{\coalescentTime{}}}  \condprob{\coEPS{\coalescentTime{}}}{\coCovariatesMatrix}.
\end{align}
Because the posterior distribution is complex and cannot be computed directly, we need to draw samples from it using a simulation method. 
A common choice is using Markov chain Monte Carlo (MCMC). 
While traditional MCMC methods, like random walk MCMC, have long been standard for sampling trees and rate parameters, they are often inefficient for sampling a large number of highly correlated parameters \citep{neal_mcmc_2011, gelman_bayesian_2013}, as the piecewise-constant population sizes often are. 

\citet{baele_hamiltonian_2020} proposed to adopt a Hamiltonian Monte Carlo (HMC) sampler to improve efficiency in sampling the effective population size levels in the context of the log-linear model  introduced by \citet{gill_understanding_2016}. 
HMC is a gradient-based MCMC sampler that is particularly well-suited for high-dimensional and highly correlated posteriors \citep{neal_mcmc_2011}. 
By using gradient information to guide the sampling process, HMC can propose larger, more efficient moves through the parameter space, overcoming the limitations of random walk methods.
Here we show how their framework can be easily extended to sample from the posterior of our proposed GP-based model.
In the following sections, we will first derive the two crucial ingredients of the HMC sampler: the (log-) coalescent-likelihood and prior's gradients and Hessians.
We will then provide a detailed explanation of how the HMC algorithm uses these quantities to efficiently sample from the posterior distribution.

\subsubsection{Gradient and Hessian Computations} \label{NPco.sec:likelihoodGradientHessian}
We begin by deriving the gradient and Hessian for the logarithm of the coalescent likelihood $\condprob{\phylogenies{}}{\coEPS{\coalescentTime{}}}$ and its corresponding prior $\condprob{\coEPS{\coalescentTime{}}}{\coCovariatesMatrix}$. 

\paragraph{Coalescent likelihood} As explained in Section~\ref{NPco.sec:coalescentBasedLikelihood}, the coalescent-based log-likelihood can be written as a sum of a distribution over the coalescent times plus a normalizing constant, that is, under a heterochronous data framework:
\begin{align}
    \log\condprob{\phylogenies{}}{\coEPS{\coalescentTime{}}} &= \log \condprob{\coalescentTimeVector{}}{\piecewiseConstantAuxiliaryVector} + \mathrm{const} \\
    &=   - \sum_{\coGridPointIndex=1}^{\conGridPoints} \conCoalescentEventsGridPoints{\coGridPointIndex} \piecewiseConstantAuxiliaryVariable{\coGridPointIndex}
    -  \sum_{\coGridPointIndex=1}^{\conGridPoints} e^{-\piecewiseConstantAuxiliaryVariable{\coGridPointIndex}} \sum_{\coTimeIndex=1}^{\conCoalescentEventsGridPoints{\coGridPointIndex}+\conSamplingTipsDatesGridPoints{\coGridPointIndex}} \binom{\conLineagesGridPoints{\coGridPointIndex ,\coTimeIndex}}{2} (\coalescentTime{\coGridPointIndex, \coTimeIndex} - \coalescentTime{\coGridPointIndex, \coTimeIndex-1}),
\end{align}
where in the second equality we have inserted Equation~\eqref{NPco.eq:coheteroPiecewiseLikelihood} substituting $\coPiecewiseConstantLevelsVector$ with the log-effective population size levels $\piecewiseConstantAuxiliaryVector$, and we have noticed that the constant cancels out. 

Applying simple differentiation rules we can derive the gradient with respect to each parameter $\piecewiseConstantAuxiliaryVariable{\coGridPointIndex}$, that is, the vector $\nabla_{\piecewiseConstantAuxiliaryVector}\log \condprob{\coalescentTimeVector{}}{\piecewiseConstantAuxiliaryVector}$ with $\coGridPointIndex$-th entry equal to:
\begin{align} \label{co.eq:likelihood-grad}
\myPartial{\log \condprob{\coalescentTimeVector{}}{\piecewiseConstantAuxiliaryVector}}{\piecewiseConstantAuxiliaryVariable{\coGridPointIndex}} = 
    -  \conCoalescentEventsGridPoints{\coGridPointIndex}
    + e^{-\piecewiseConstantAuxiliaryVariable{\coGridPointIndex}} \sum_{\coTimeIndex=1}^{\conCoalescentEventsGridPoints{\coGridPointIndex}+\conSamplingTipsDatesGridPoints{\coGridPointIndex}} \binom{\conLineagesGridPoints{\coGridPointIndex ,\coTimeIndex}}{2} (   \coalescentTime{\coGridPointIndex, \coTimeIndex} - \coalescentTime{\coGridPointIndex, \coTimeIndex-1}).
\end{align}
Besides, the Hessian $\nabla_{\piecewiseConstantAuxiliaryVector}^2\log \condprob{\coalescentTimeVector{}}{\piecewiseConstantAuxiliaryVector}$ is a diagonal matrix with diagonal elements: 
\begin{align} \label{NPCo.eq:likelihood-hessian}
\mySecondPartial{\log \condprob{\coalescentTimeVector{}}{\piecewiseConstantAuxiliaryVector}}{\piecewiseConstantAuxiliaryVariable{\coGridPointIndex}} = 
    - e^{-\piecewiseConstantAuxiliaryVariable{\coGridPointIndex}} \sum_{\coTimeIndex=1}^{\conCoalescentEventsGridPoints{\coGridPointIndex}+\conSamplingTipsDatesGridPoints{\coGridPointIndex}} \binom{\conLineagesGridPoints{\coGridPointIndex, \coTimeIndex}}{2}  (   \coalescentTime{\coGridPointIndex, \coTimeIndex} - \coalescentTime{\coGridPointIndex, \coTimeIndex-1}).
\end{align}
\paragraph{Prior gradient and Hessians} From Equation~\eqref{co.eq:gmrfCentricLogPopDistribution}, we can write the log-prior of $\piecewiseConstantAuxiliaryVector$ conditional on the mean vector $\coGmrfMeanVector{}{\coCovariatesMatrix}$ as:
\begin{equation}\label{eq:gmrf-logprior-x}
\log \prob{\piecewiseConstantAuxiliaryVector \mid \coGmrfMeanVector{}{\coCovariatesMatrix},\coGmrfPrecision}
= -\frac{\coGmrfPrecision}{2}\, \left[ \piecewiseConstantAuxiliaryVector-\coGmrfMeanVector{}{\coCovariatesMatrix} \right]^\top
\coGmrfPrecisionMatrix\,
\left[ \piecewiseConstantAuxiliaryVector-\coGmrfMeanVector{}{\coCovariatesMatrix} \right]  + \mathrm{const}.
\end{equation}
Hence the gradient and (negative-definite) Hessian w.r.t.\ $\piecewiseConstantAuxiliaryVector$ are
\begin{align}
\nabla_{\piecewiseConstantAuxiliaryVector}\log \prob{\piecewiseConstantAuxiliaryVector \mid \coGmrfMeanVector{}{\coCovariatesMatrix},\coGmrfPrecision}
&= -\coGmrfPrecision\,\coGmrfPrecisionMatrix \left[ \piecewiseConstantAuxiliaryVector-\coGmrfMeanVector{}{\coCovariatesMatrix}\right], \label{eq:prior-grad-x}\\
\nabla^2_{\piecewiseConstantAuxiliaryVector}\log \prob{\piecewiseConstantAuxiliaryVector \mid \coGmrfMeanVector{}{\coCovariatesMatrix},\coGmrfPrecision}
&= -\coGmrfPrecision\,\coGmrfPrecisionMatrix. \label{eq:prior-hess-x}
\end{align}


\paragraph{Posterior gradient and Hessian.}
The posterior gradient $\nabla_{\piecewiseConstantAuxiliaryVector}\log \condprob{\piecewiseConstantAuxiliaryVector}{\coalescentTimeVector{},\coCovariatesMatrix}$ and Hessian $\nabla_{\piecewiseConstantAuxiliaryVector}^2\log \condprob{\piecewiseConstantAuxiliaryVector} {\coalescentTimeVector{},\coCovariatesMatrix}$ follow directly from the sum of the corresponding prior and likelihood contributions.

\subsubsection{Sampling with HMC}  \label{NPco.sec:HMCSampling}
The existence of a computationally feasible exact gradient for the coalescent likelihood and for the GP prior enables efficient inference under the Bayesian framework using an HMC sampler \citep{neal_mcmc_2011}. 
Given a parameter of interest $\coHMCsamplerParameter$ with posterior distribution $\coHMCposteriorDistribution{\coHMCsamplerParameter}$, the \textbf{HMC sampler} provides a Metropolis-Hastings proposal \citep{metropolis_equation_1953} by augmenting the sampling space with an auxiliary variable, and exploring this augmented space simulating the Hamiltonian dynamics. 
In the language of these dynamics, the parameter of interest $\coHMCsamplerParameter$ is referred to as the ``location'' component of the system, whereas the auxiliary variable $\coHMCsamplerMomentum$ is interpreted as the ``momentum'' or ``velocity''. 
$\coHMCsamplerLocation$ and $\coHMCsamplerMomentum$ are assumed to be independent; the distribution of the former is the target posterior distribution $\coHMCposteriorDistribution{\coHMCsamplerLocation}$, while the latter is typically modeled as a centered multivariate normal random variable with covariance matrix $\coHMCmassMatrix$.

The system is globally described by a total energy function $\coHMChamiltonian{\coHMCsamplerParameter}{\coHMCsamplerMomentum}$, said Hamiltonian function,  defined as the sum of the ``potential energy'' $\coHMCpotentialEnergy(\coHMCsamplerParameter)$ and the ``kinetic energy'' $ \coHMCkyneticEnergy(\coHMCsamplerMomentum)$ of the system. 
In turn, the potential energy $\coHMCpotentialEnergy(\coHMCsamplerParameter)$ is set to the negative of the log-posterior density $\coHMCpotentialEnergy(\coHMCsamplerParameter) = - \log \coHMCposteriorDistribution{\coHMCsamplerParameter}$, while the kinetic energy function is equal to $\coHMCkyneticEnergy(\coHMCsamplerMomentum) = \coHMCsamplerMomentum\transpose \coHMCmassMatrix^{-1} \coHMCsamplerMomentum$. 
Starting from a current state $(\coHMCsamplerParameter_0,\coHMCsamplerMomentum_0)$, HMC generates a MH proposal \citep{metropolis_equation_1953} by simulating the Hamiltonian dynamics in the space $(\coHMCsamplerParameter, \coHMCsamplerMomentum)$, which is governed by the following differential equations:
\begin{equation}\label{NPco.eq:hmcDifferentialEquations}
\begin{aligned}
    \frac{d \coHMCsamplerParameter}{d\coHMCtimeValue} &= \;\,\,\, \nabla \coHMCkyneticEnergy(\coHMCsamplerMomentum) = \coHMCmassMatrix^{-1}\coHMCsamplerMomentum  \\
    \frac{d \coHMCsamplerMomentum}{d\coHMCtimeValue} &= - \nabla \coHMCpotentialEnergy(\coHMCsamplerParameter) = \nabla \log\coHMCposteriorDistribution{\coHMCsamplerParameter}.
\end{aligned}
\end{equation}
One successful way to numerically approximate these dynamics is based on the \textit{leapfrog algorithm} \citep{neal_mcmc_2011} that integrates the dynamics forward in time through discrete steps of size $\coHMCleapfrogStep$ according to: 
\begin{align}
    \coHMCsamplerMomentum_{\coHMCtimeValue + \frac{\coHMCleapfrogStep}{2}} 
    &= \coHMCsamplerMomentum_{\coHMCtimeValue} + \frac{\coHMCleapfrogStep}{2}\nabla \log\coHMCposteriorDistribution{\coHMCsamplerParameter_{\coHMCtimeValue}} \nonumber \\
        \coHMCsamplerParameter_{\coHMCtimeValue + \coHMCleapfrogStep} 
    &= \coHMCsamplerParameter_\coHMCtimeValue + \coHMCleapfrogStep \coHMCmassMatrix^{-1}\coHMCsamplerMomentum_{\coHMCtimeValue + \frac{\coHMCleapfrogStep}{2}} \\
    \coHMCsamplerMomentum_{\coHMCtimeValue + \coHMCleapfrogStep} 
    &= \coHMCsamplerMomentum_{\coHMCtimeValue + \frac{\coHMCleapfrogStep}{2}}  + \frac{\coHMCleapfrogStep}{2}\nabla \log\coHMCposteriorDistribution{\coHMCsamplerParameter_{\coHMCtimeValue+ \coHMCleapfrogStep}}. \nonumber
\end{align}
Simulating the Hamiltonian for a time $\coHMCtimeValue$ requires a total of $\coHMCtimeValue / \coHMCleapfrogStep$ steps, which in turn require about the same number of gradient evaluations. 
However, HMC proposals tend to be rather uncorrelated with the current states and are accepted with high probability \citep{neal_mcmc_2011}. 
These features make such an algorithm particularly effective when sampling from high-dimensional correlated distributions \citep{beskos_optimal_2013, gelman_bayesian_2013}.

\paragraph{Preconditioning} 
A common choice for the covariance matrix $\coHMCmassMatrix$ is the identity matrix. 
A major motivation is that this approach avoids a costly matrix inversion (cubic time complexity for dense unstructured matrices) at each leapfrog step.
However, the efficiency of the HMC sampler can substantially benefit by leveraging the geometrical structure of the posterior distribution through an ``informed'' mass matrix \citep{neal_mcmc_2011}: a technique known as \textbf{preconditioning}. 
A candidate choice is to set the mass matrix equal to the negative of the posterior Hessian matrix \citep{girolami_riemann_2011}. 
Fortunately, in the current context, the Hessian is a tri-diagonal matrix and so computing the product of its inverse with the momentum, $\coHMCmassMatrix^{-1}\coHMCsamplerMomentum$, is an operation with linear time complexity using a specialized Gaussian elimination algorithm.
An even faster alternative is to set all the non-diagonal elements directly equal to zero. This implies that at each leapfrog step the only preconditioning-related operation is the scaling of the momentum variables by the inverse of each of the Hessian diagonal elements.




\section{Results}\label{NPco.sec:results}
In this section, we validate our model using two synthetic and three real data examples. 
Specifically, we will first simulate and analyze two trees where we fix the logarithm of effective population sizes to linear and nonlinear functions of a covariate. 
Second, we study the effect of temperature fluctuations on the demographic history of the yellow fever virus (YFV) in Brazil based on samples collected between $2016$ and $2018$ \citep{hill_climate_2022}, and of the late Quaternary musk ox globally \citep{campos_ancient_2010, gill_understanding_2016}. 
Then, we leverage the local incidence rate of HIV to track the population dynamics of the HIV-1 CRF02\_AG Clade in Cameroon between $1996$ to $2004$ \citep{brennan_prevalence_2008,faria_phylodynamicsHIV_2012}. 
All the following analyses are run using the publicly available software BEAST X \citep{baele_beast_2025} supported by the high-performance computing environment BEAGLE  \citep{ayres_beagle_2019}. 
The code and data to reproduce the analyses are available in a public repository (\url{https://github.com/suchard-group/NonparametricCoalescentProcesses}).

\subsection{Simulations} \label{NPco.sec:simulation}
\begin{figure}[!htbp]
    \centering
    \includegraphics[scale=1]{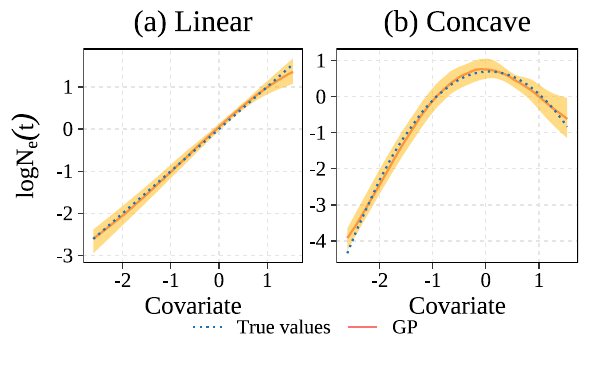} 
    \caption{Simulations: $\operatorname{log}\coEPS{\coalescentTime{}}$ vs Covariates}
    \label{NPco.fig:simulations}
    \vspace{5pt} 
    \justify
    \footnotesize \textbf{Note}: The two plots depict the relationship between the log effective population sizes and a covariate under two simulation scenarios. 
    The chosen covariate is the average temperature in Brazil between $2014$ and $2019$ \citep{hill_climate_2022}. 
    Two trees were simulated under a piecewise-constant coalescent model, where the $\operatorname{log}\coEPS{\coalescentTime{}}$ levels were set respectively equal to a linear (left plot) and a nonlinear (right) function of the covariate, as shown by the two dotted lines. 
    The solid lines represent the median estimate of the two functions under a GP model, while the shaded areas show the $95\%$ HPD intervals.
\end{figure}

To assess the performance of the proposed methods in estimating the model parameters, we conduct two simulation studies. 
Our reference data is based on the findings of \citet{hill_climate_2022}, who studied the relationship between the average temperature areas with yellow fever virus circulation across Brazil and the virus' demographic fluctuations from 2016 to 2018, based on $705$ local samples.
Subsequently, we simulate two phylogenetic trees, each with $705$ taxa, under a piecewise-constant coalescent model. 
For these simulations, we set the log effective population sizes equal to, respectively, a linear and a concave function of the (standardized) average temperature values from \citet{hill_climate_2022}. 
These two functions are represented by the dashed lines in the two plots of Figure \ref{NPco.fig:simulations}.

We then analyze the two simulated trees using a GP prior with temperature as a covariate. 
We adopt a squared exponential kernel and place an exponential prior with rate one on its two hyperparameters.
The two plots in Figure \ref{NPco.fig:simulations} present the median $\operatorname{log}\coEPS{\coalescentTime{}}$ (solid lines) as a function of the covariate, along with the corresponding $95\%$ highest posterior density (HPD) intervals, for both the linear (left plot) and the concave (right plot) cases. 
The close alignment between the inferred $\operatorname{log}\coEPS{\coalescentTime{}}$ and the true values demonstrates the validity and accuracy of our methods.

\begin{figure}[!htbp]
    \centering
    \includegraphics[width=\linewidth]{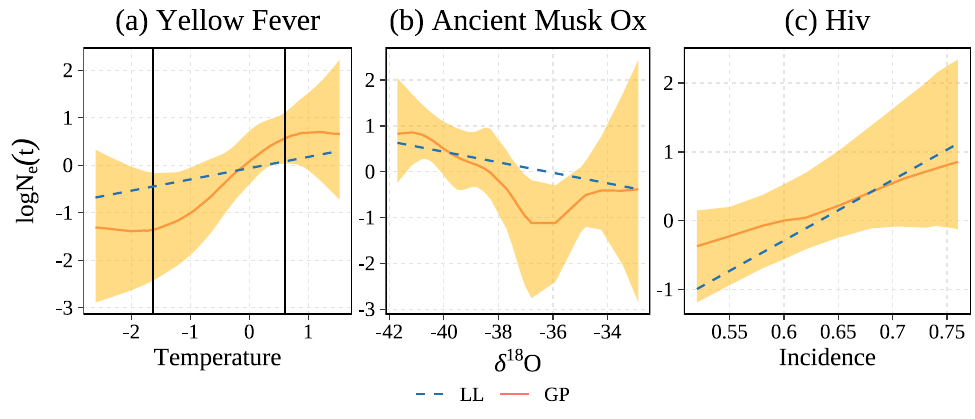} 
    \caption{Data Examples: $\operatorname{log}\coEPS{\coalescentTime{}}$ vs Covariates}
    \label{NPco.fig:logPopVsCovariates}
    \justify
    \footnotesize \textbf{Note}: The three plots compare the relationship between the log effective population sizes and three covariates - temperature, $\delta^{18} O$ (proxy for inverse temperature), and (HIV) incidence rate - using a log-linear (LL) and a Gaussian process (GP) model respectively in three data examples focused on: (a) yellow fever virus in Brazil ($2016$-$18$) from \citet{hill_climate_2022}, (b) ancient musk ox  \citep{campos_ancient_2010, gill_understanding_2016}, (c) and the CRF02\_AG strain of HIV in Cameroon ($1990-2004$) \citep{brennan_prevalence_2008, gill_understanding_2016}. 
    Within the inferential process, the evolutionary trees are assumed to be random, and they are modeled using a coalescent prior with piecewise-constant $\coEPS{\coalescentTime{}}$.
    The solid lines represent the median estimate of the $\operatorname{log}\coEPS{\coalescentTime{}}$ levels as functions of the covariates under a GP model, and the shaded areas show the corresponding $95\%$ HPD intervals. 
    The dashed lines represent the median $\operatorname{log}\coEPS{\coalescentTime{}}$ under the LL model. Finally, the black vertical lines in the first figure represent the ``flattening'' points, that is, when the temperature effect changes sign or becomes null.
\end{figure}

\subsection{Temperature Effect on Yellow Fever Virus Demography in Brazil} \label{NPco.sec:yellowFever}
After having validated our method on two simulated datasets, we apply it to real data. 
We start again from \citet{hill_climate_2022}, where the authors analyzed the relationship between climate and demography of the yellow fever virus (YFV) between 2014 and 2018 in Southeast Brazil. 
Among other findings, they detected a positive relationship between YFV population size and temperature levels, suggesting that higher temperatures facilitate the reproduction and dispersion of the virus. 
To achieve this, the authors modeled the piecewise-constant effective population sizes of YFV as a log-linear function of the average local temperatures in Southeast Brazil. 
Here, we aim at relaxing the restrictive log-linear assumption to both better inform the inferential process and detect potential nonlinearities in the temperature effect.
To accomplish this, we perform a phylogenetic analysis consistent with \citet{hill_climate_2022} using $705$ nearly-complete YFV genomes sampled in humans, neotropical primates, and mosquitoes. 
However, we model the piecewise-constant effective population sizes using a flexible GP prior with temperature as a covariate. 
As in Section \ref{NPco.sec:simulation}, we adopt a squared exponential kernel and place an exponential prior with rate $1$ on its two hyperparameters.

Figure \ref{NPco.fig:logPopVsCovariates}(a) depicts the median $\operatorname{log}\coEPS{\coalescentTime{}}$ as a function of temperature levels obtained, respectively, using a log-linear (dashed line) and a GP (solid line) model. 
The GP-based function reveals a sigmoidal nonlinear relationship between population size and temperature. 
In particular, we identify two ``flattening'' points represented by the vertical black lines in Figure~\ref{NPco.fig:logPopVsCovariates}(a): when the temperature has a value between those two points, an increase in temperature is associated with demographic growth; conversely, an increase in temperature at values outside that interval has no or negligible effect on the population (the derivative is less than $0.05$).
Since the log-linear model assumes a constant temperature effect, it interpolates the nonlinear function and so it respectively under- and over-estimates the impact of an increase in temperature within and outside the central range. 

\begin{figure}[!htbp]
    \centering
    \includegraphics[width=\linewidth]{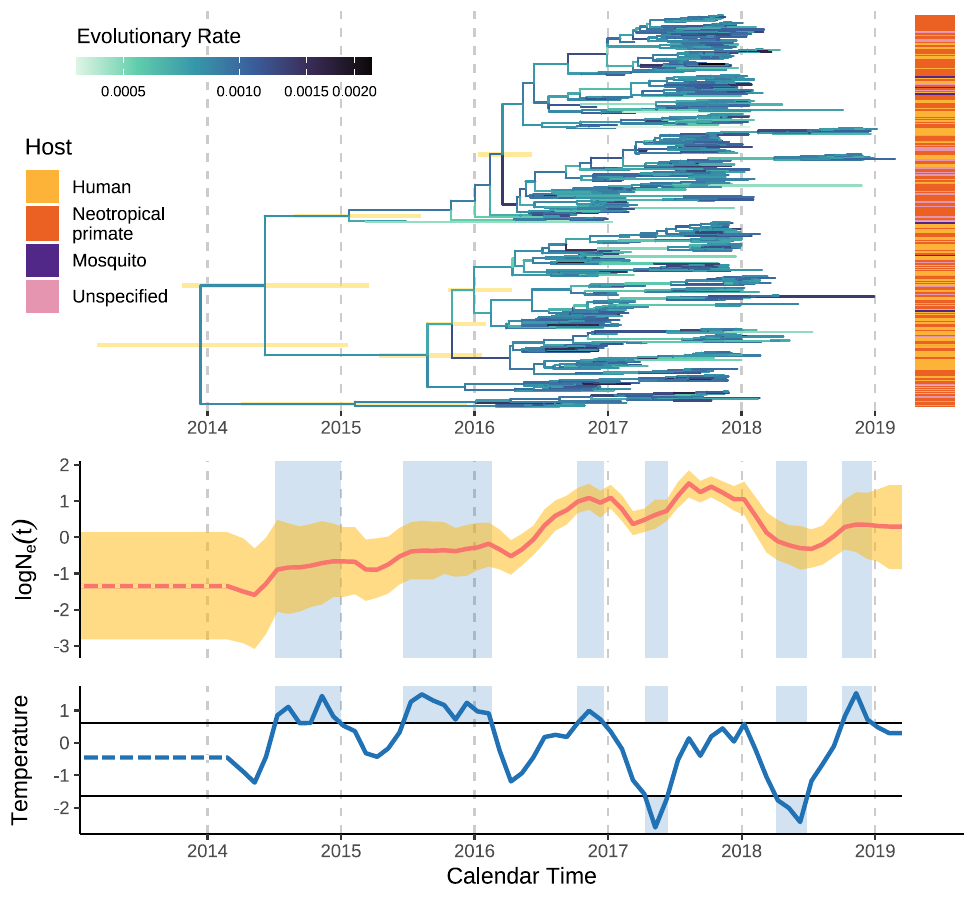} 
    \caption{Yellow Fever Virus: Tree and demography}
    \label{NPco.fig:YFtree}
    \vspace{5pt} 
    \justify
    \footnotesize \textbf{Note}: The upper part of the plot depicts the evolutionary tree of the yellow fever virus (YFV) in Brazil based on samples collected between $2016$ and $2018$ in humans, neotropical primates, mosquitoes, and other hosts as specified by the colors of the vertical rectangle on the right \citep{hill_climate_2022}.
    The shaded yellow horizontal rectangles represent the $95\%$ HPD intervals for some node ages, while the branches are colored based on their inferred branch-constant DNA substitution evolutionary rate. 
    The two line plots in the lower part depict the trends of YFV $\operatorname{log}\coEPS{\coalescentTime{}}$ levels and the average local temperature in Brazil between $2014$ and $2019$. 
    The red line shows the median $\operatorname{log}\coEPS{\coalescentTime{}}$ inferred under a GP model, while the yellow shaded area represents the corresponding $95\%$ HPD intervals. 
    The shaded blue vertical bands correspond to time frames where the temperature is respectively higher or lower than the two ``flattening'' points represented here by the two horizontal black lines, and in Figure \ref{NPco.fig:logPopVsCovariates}(a) by the two vertical lines. 
    While in the white areas, an increase in temperature is linked to YFV demographic growth, in the shaded areas, it is associated with demographic stability.
    Before $2014$ the temperature has been assumed to be constant (dashed line) and equal to the earliest available value from \citet{hill_climate_2022}.
\end{figure}

In the two line plots at the bottom of Figure~\ref{NPco.fig:YFtree} we compare the trends of temperature and $\coEPS{\coalescentTime{}}$ over time.
The vertical shaded bands are drawn in correspondence with temperatures respectively lower and higher than the aforementioned flattening points. 
A close inspection reveals that temperature and $\coEPS{\coalescentTime{}}$ tend to evolve in parallel at temperatures in the range between the two flattening points (white area), while they look rather independent outside it.

\subsection{Population Dynamics of Late Quaternary Musk Ox} \label{NPco.sec:ancientMox}
The decline and extinction of large mammals during the Late Quaternary period are well-documented, but the causes are still debated \citep{lorenzen_species-specific_2011}. 
Most of the discussion centers on whether climate change or human activity was the primary factor \citep{lorenzen_species-specific_2011}. 
Researchers often use ancient genetic material to reconstruct population history, which can then be compared to climate and fossil records to shed light on this debate.

A 2010 study by \citet{campos_ancient_2010} used ancient DNA to reconstruct the population history of musk oxen based on Bayesian Skygrid \citep{minin_smooth_2008} and Skyline \citep{drummond_bayesian_2005} models.
They found that the musk ox population did not seem to be affected by the arrival of humans in their territory. 
However, the study did find a correlation between population growth and cooler climate periods, with populations declining during warmer, more unstable times. 
This suggested that environmental changes, not humans, were the main cause of the musk ox's population changes. 
To further investigate this, \citet{gill_understanding_2016} enriched the Skygrid with a log-linear model based on a proxy for temperature typically used to reconstruct ancient climate trends: oxygen isotope records. 
As a measure of oxygen isotope composition, they adopt the ice core $\delta^{18} O$ data from the Greenland Ice Core Project \citep{dansgaard_evidence_1993, grootes_comparison_1993} averaged across intervals of $3000$ years, where higher levels of $\delta^{18} O$ correspond to warmer polar temperatures. 
The authors' inferred demographic history suggests an inverse relationship between climate and $\delta^{18} O$ levels, confirming the findings in \citet{campos_ancient_2010}. 
However, the negative covariate effect size in the log-linear model does not reveal a statistically significant association between the two variables.

Here, we investigate whether the log-linear assumption was too restrictive to capture the relationship. We reproduce the phylogenetic analysis using the same data as in \citet{campos_ancient_2010} and \citet{gill_understanding_2016}. 
In particular, we use sequence data composed of $682$ bp of the mitochondrial control region sampled from $149$ specimens dated between present and 56,900 radiocarbon years before present (YBP). 
The samples were collected in
the Taimyr Peninsula ($n = 54$), the Urals ($n = 26$), Northeast Siberia ($n = 12$), North America ($n = 14$), and Greenland ($n = 43$). 

\begin{figure}[!htbp]
    \centering
    \includegraphics[width=\linewidth]{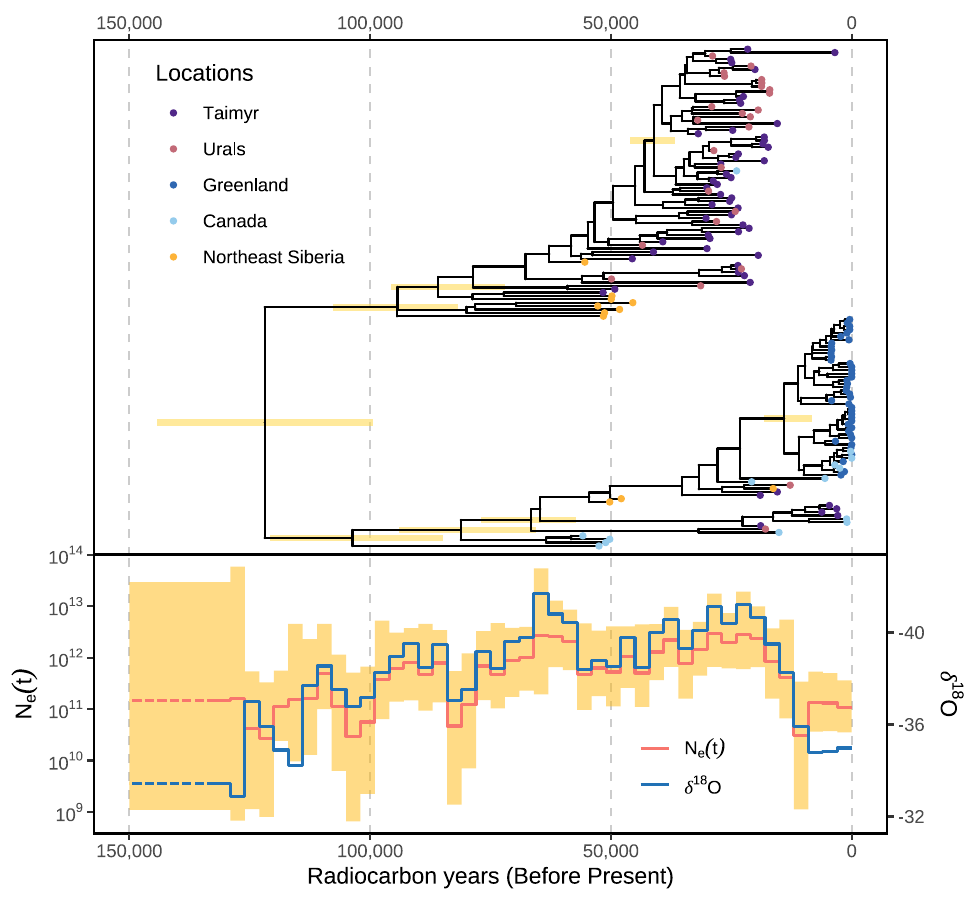} 
    \caption{Ancient Musk Ox: Tree and Demography}
    \label{NPco.fig:MoxTree}
    \vspace{5pt} 
    \justify
    \footnotesize \textbf{Note}: The upper part of the plot illustrates the evolutionary tree of the late Quaternary musk oxen using specimens dated between present and 56,900 radiocarbon years before present (YBP). 
    The tip colors represent the sample locations, while the shaded yellow horizontal rectangles show the $95\%$ HPD intervals for some node ages.
    The bottom line plot depicts the trends of median ancient musk oxen $\coEPS{\coalescentTime{}}$ levels (red line) learned under the GP model, and ice core $\delta^{18} O$ (blue line) levels from the Greenland Ice Core Project \citep{dansgaard_evidence_1993, grootes_comparison_1993}. 
    $\delta^{18} O$ is a measure of oxygen isotope composition, and it is used as a proxy for temperature \citep{gill_understanding_2016}. 
    Therefore, the plot suggests an \textit{inverse} relationship between temperature and demographic growth (note that the covariate axis is decreasing to accentuate the correlation).
    The yellow shaded area represents the $95\%$ HPD intervals for $\coEPS{\coalescentTime{}}$.
    On the left side, the lines are dashed since the $\delta^{18} O$ level was assumed to be constant and equal to the earliest available value from \citet{gill_understanding_2016}.
\end{figure}

Using a GP model with a squared exponential kernel and rate-$1$ exponential priors for the hyperparameters, we also detect an inverse relationship between $\delta^{18} O$ and musk ox log effective population sizes (Figure~\ref{NPco.fig:logPopVsCovariates}(b)). 
We observe that at lower temperatures ($\delta^{18} O < -37$), the effect of warming appears higher than what the LL model suggests, that is, the corresponding population size contraction is higher. 
Besides, at higher temperatures ($\delta^{18} O \geq -37$), the effect size appears close to $0$ with a substantial uncertainty, most probably due to the limited sample size. 
Figure \ref{NPco.fig:MoxTree} shows the trends of $\coEPS{\coalescentTime{}}$ (red line) and $\delta^{18} O$ (blue line) over time. 
We have plotted $\delta^{18} O$ using a decreasing axis to make the correlation evident; therefore, the fact that the two lines are overall parallel means temperature and $\coEPS{\coalescentTime{}}$ are \textit{inversely} related. 
By contrast, we notice that at higher temperatures the relationship looks, if anything, positive, even though there is substantial uncertainty as the wide HPD intervals suggest. 
Notably, the GP-based approach provides a nuanced quantification of how uncertainty in the temperature effect varies across different temperature levels, supporting the idea that the effect of a warming climate is more certain at lower temperatures and considerably more uncertain at higher temperatures.

Finally, above the demographic plot in Figure \ref{NPco.fig:MoxTree}, we present the maximum clade credibility (MCC) tree describing the evolutionary history of ancient musk ox, where the tips are colored by the sample locations. 
We identify two major clades that mostly separate the samples collected in Greenland and Canada from those collected in the Taimyr Peninsula and the Urals, consistent with previous work.

\subsection{Demographic History of the HIV-1 CRF02\_AG Clade in Cameroon} \label{NPco.sec:HIV}
Recombinant HIV-1 genomes, known as circulating recombinant forms (CRFs), are created when two or more different HIV-1 subtypes combine. 
CRF02\_AG, a specific type of HIV-1 CRF, causes only a small percentage of global HIV infections ($7.7\%$), but it is responsible for the majority ($60–70\%$) of cases in Cameroon \citep{powell_evolution_2010}.
Here, we study the population history of this HIV strain in Cameroon using a multilocus alignment of 336 \textit{gag}, \textit{pol}, and \textit{env} CRF02\_AG gene sequences from 336 HIV samples collected from blood donors in Yaounde and Douala between 1996 and 2004 \citep{brennan_prevalence_2008}. 
Our work builds on previous research by \citet{faria_phylodynamicsHIV_2012} and  \citet{gill_improving_2013} where the authors adopted a coalescent-based piecewise-constant model for the HIV population sizes, and on its extension in \citet{gill_understanding_2016}, where the inferential process was supported by a log-linear model using yearly HIV incidence rates for adults (ages 18–49s) in Cameroon as a covariate. 

This example is of particular interest for two reasons: first, there are three separate genes, meaning that we fall under the multilocus framework introduced in Section \ref{NPco.sec:coalescentBasedLikelihood}; second, we are in a scenario with missing covariate values, since incidence rates were available only after $1990$. 
We handle the missing pre-$1990$ incidence rates by assigning them a GMRF prior anchored at the earliest observed value. 
We then sample from the posterior distribution of the missing covariates with HMC.

\begin{figure}[!htbp]
    \centering
    \includegraphics[width=\linewidth]{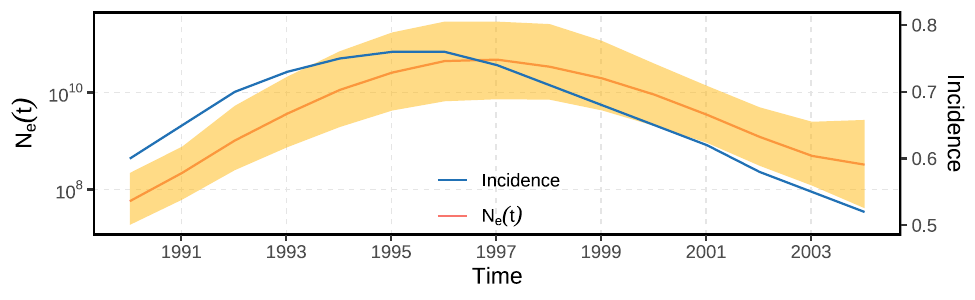} 
    \caption{HIV Example: Demography}
    \label{NPco.fig:HIV}
    \justify
    \footnotesize \textbf{Note}: The line plot depicts the trends of median effective population size levels for the HIV CRF02\_AG strain in Cameroon (red line) learned under the GP model, and the overall local HIV incidence rate (blue line). 
    These $\coEPS{\coalescentTime{}}$ levels were inferred using a Gaussian process (GP) model, with the incidence rate incorporated as a covariate.
    Inference was performed assuming a random tree under a coalescent prior with piecewise-constant $\coEPS{\coalescentTime{}}$ modeled using a GP using the incidence rate as a covariate.
    The yellow shaded area represents the $95\%$ HPD intervals for the $\coEPS{\coalescentTime{}}$ levels.
    The visual trends in the plot suggest a direct relationship between the incidence rate and the $\coEPS{\coalescentTime{}}$ levels.
\end{figure}

Using a more flexible, incidence-rate-informed GP model yields results consistent with prior work.
In particular, the effective population size of this HIV strain increased until reaching a peak in 1997, after which it declined steadily. (Figure \ref{NPco.fig:HIV}). 
Our analysis also confirms the strong link between the population history of CRF02\_AG and the incidence of HIV found in \citet{gill_understanding_2016}, specifically supporting the validity of the log-linear assumption (Figure \ref{NPco.fig:logPopVsCovariates}(c)). 
This relationship is also evident from Figure \ref{NPco.fig:HIV}, which shows that the inferred population sizes and incidence rates follow a similar pattern over time.

\section{Discussion}
We have developed a flexible framework for coalescent-based inference that replaces log-linear covariate models with Gaussian process priors, enabling the detection of complex nonlinear relationships between covariates and effective population size. 
Our approach addresses a key limitation in existing methods: the assumption that covariates affect demography in a strictly (log-) linear manner. 
When this assumption is violated, a linear model fails to properly leverage the information contained in the covariates to improve overall inference, and produces a biased estimate of their effects.

\paragraph{Empirical insights and method performance.} We showcase our method on three empirical applications, demonstrating the advantages brought about by its flexibility.
In the yellow fever virus analysis, we uncovered a sigmoidal relationship between temperature and viral demography, with two "flattening points" that delineate a temperature window where warming drives population growth. Outside this range, temperature increases have negligible demographic effects---a pattern the log-linear model systematically mischaracterizes by over- or under-estimating effects at different temperature levels. 
Importantly, our findings broadly align with the fact that mosquito-borne virus transmission is strongly temperature-dependent and follows a unimodal pattern: transmission suitability increases with temperature up to an optimal range and then declines at higher temperatures due to reduced mosquito survival and physiological stress \citep{Bellone:2020aa,desouza_effectsClimateChange_2024, abreu2025roleclimatechange}.

For ancient musk ox, the GP-based model revealed differential uncertainty in the temperature effect magnitude across temperature values: the inverse relationship between warming and population size is well-supported at lower temperatures but highly uncertain at higher values, a nuance obscured by the rigid stochastic structure of linear models. Conversely, our HIV-1 analysis confirmed a log-linear relationship between population sizes and incidence levels. When the true relationship is approximately linear, the GP-based model will recover it, demonstrating that it generalizes well to its linear counterpart.

\paragraph{When to use flexible covariate models.} We recommend that researchers consider GP-based models when: (1) prior knowledge suggests nonlinear or threshold effects (e.g., species thermal tolerances, disease transmission optima); (2) preliminary linear models show poor fit or residual patterns; (3) covariates span wide ranges where effects may vary; or (4) uncertainty quantification across covariate values is scientifically important. 
When covariates are uninformative or relationships are demonstrably linear, simpler models remain appropriate.

\paragraph{Limitations and considerations.} Model effectiveness depends critically on covariate quality and relevance. 
A potential problem may arise when the $\coEPS{\coalescentTime{}}$ is strongly associated with a covariate during a short period of time, as already noted in \citet{gill_understanding_2016}. 
Indeed, this could cause a poor extrapolation of the covariate effect for other time periods, leading to incorrect demographic reconstruction. 
\citet{gill_understanding_2016} already explained that this impact is limited unless the sequence data for those other time periods are weakly informative. 
Our GP framework further mitigates this issue by accommodating varying uncertainty quantification along the covariate range.

\paragraph{Future directions.} This work enables several new research directions. 
Methodologically, even though we have proposed examples with only one covariate at a time, the framework naturally extends to multiple covariates, motivating development of automated covariate selection procedures. 
Incorporating measurement error in covariates through hierarchical modeling would also further improve inference. 
Our approach to missing covariates (using GMRF priors anchored to observed values, as in the HIV analysis) could be formalized and extended for this purpose. 
Scientifically, the method empowers investigations aimed at detecting complex drivers of demographic variation, such as environmental factors. 


\paragraph{Broader implications.} By unifying temporal and covariate information through GMRF and GP priors, our framework provides more accurate and interpretable reconstruction of population histories. 
The ability to detect nonlinear covariate effects, to quantify how uncertainty varies across covariate ranges, and to accommodate missing data makes this approach particularly valuable for understanding how populations respond to exogenous drivers such as environmental fluctuations—a question of central importance in phylodynamics, disease ecology, and conservation.

\section*{Acknowledgments} We gratefully acknowledge support from Advanced Micro Devices, Inc.~with the donation of parallel computing resources used for this research. 

\section*{Funding}

{\sloppy
This work was supported by National Institutes of Health grants
U19~AI135995, R01~AI153044, R01~AI162611, and R01~AI192139.
PL acknowledges support from the Research Foundation--Flanders
(\emph{Fonds voor Wetenschappelijk Onderzoek--Vlaanderen};
G0D5117N, G0B9317N, and G051322N).
MUGK acknowledges funding from The Rockefeller Foundation (PC-2022-POP-005), the Health AI Programme from Google.org, the Oxford Martin
School Programmes in Pandemic Genomics \& Digital Pandemic Preparedness,
the European Union’s Horizon Europe programme projects MOOD (No.~874850)
and E4Warning (No.~101086640), Wellcome Trust grants
303666/Z/23/Z, 226052/Z/22/Z, and 228186/Z/23/Z,
United Kingdom Research and Innovation (No.~APP8583),
the Medical Research Foundation (MRF-RG-ICCH-2022-100069),
UK International Development (No.~301542-403),
the Bill \& Melinda Gates Foundation (INV\-063472 and INV\-090281),
and the Novo Nordisk Foundation (NNF24OC\-0094346).
NRF acknowledges support from the Medical Research Council--São Paulo Research
Foundation (FAPESP) CADDE partnership award (MR/S0195/1 and FAPESP~18/14389-0),
the Wellcome Trust Digital Technology Development Award in Climate Sensitive
Infectious Disease Modelling (226075/Z/22/Z), the Wellcome Trust DeZi Network programme (Dengue and Zika Immunology and Genomics Multi-Country Network;
316633/Z/24/Z), Temasek Foundation (UNITEDengue), the MRC Centre for Global
Infectious Disease Analysis (MR/R015600/1), jointly funded by the UK Medical
Research Council (MRC) and the UK Foreign, Commonwealth \& Development Office
(FCDO) under the MRC/FCDO Concordat agreement, and as part of the EDCTP2
programme supported by the European Union.
The contents of this publication are the sole responsibility of the authors and do not
necessarily reflect the views of the European Commission or the other funders.
}


\section*{Data Availability}
All data, code, and analysis files are available from GitHub 
(\url{https://github.com/suchard-group/NonparametricCoalescentProcesses}). 
The repository includes sequence alignments, covariate data, BEAST X configuration files, analysis scripts, and results for all examples presented in this study.

\section*{Conflict of Interest}
The authors declare no conflicts of interest.

\bibliography{references}
\end{document}